# High temperature Néel skyrmions in simple ferromagnets

Peng Wang[1,6], Rana Saha[1,5,6], Holger L. Meyerheim[1], Ke Gu[1], Hakan Deniz[1], David Eilmsteiner[2], Andrea Migliorini[1], Juan Rubio Zuazo[3,4], Engenia Sebastiani-Tofano[3,4], Ilya Kostanovski[1], Abhay Kant Srivastava[1], Arthur Ernst[1,2], Stuart S. P. Parkin[1*]

[1]Max Planck Institute of Microstructure Physics, Weinberg 2, 06120 Halle (Saale), Germany

[2]Institute for Theoretical Physics, Johannes Kepler University Linz, Altenberger Strasse 69, A-4040, Linz, Austria

[3]Spanish CRG Beamline BM25-SpLine at the ESRF, 38043 Grenoble, France

[4]Instituto de Ciencia de Materiales de Madrid-CSIC, 28049, Madrid, Spain

[5]Department of Chemistry, Indian Institute of Science Education and Research, Yerpedu, Tirupati 517619, India

[6]These authors contribute equally.

*e-mail: stuart.parkin@mpi-halle.mpg.de

**A wide variety of chiral non-collinear spin textures have been discovered and have unique properties that make them highly interesting for technological applications. However, many of these are found in complex materials and only in a narrow window of temperature. Here, we show the formation of Néel-type skyrmions in thin layers of simple ferromagnetic alloys, namely Co-Al and Co-Ni-Al, over a wide range of temperature up to ~770 K, by imposing a vertical strain gradient via epitaxy with an Ir-Al underlayer. The Néel skyrmions are directly observed using Lorentz transmission electron microscopy in freestanding membranes at high temperatures and the strain gradient is directly measured from x-ray diffraction anomalous peak profiles. Our concept allows simple centrosymmetric ferromagnets with high magnetic ordering temperatures to exhibit "hot" skyrmions, thereby, bringing closer skyrmionic electronics.**

Recently, a zoology of chiral spin textures have been discovered in a wide range of ferro- and ferri-magnetic compounds typically in a narrow window of temperature[1,2]. These spin textures include Bloch[3] and Néel[4,5] skyrmions, perhaps the simplest of these spin textures, as well as more complex textures such as anti-skyrmions[6] and many more[7,8]. Each of these are typically nanoscopic in size, ranging from tens to hundreds of nanometers[9,10], and exhibit many shapes and forms[7,11]. Beyond their fundamental interest, their application as non-volatile bits for non-volatile memory and computing devices has been proposed[12,13]. One of the biggest challenges to date is the stabilization of skyrmions in thin films well above ambient temperature. A prerequisite for the formation of many of these spin textures is a vector Dzyaloshinskii-Moriya magnetic exchange Interaction (DMI)[14,15] which is only possible in bulk compounds without crystal inversion symmetry[16,17] or at interfaces[18] which considerably restricts the number of suitable magnetic materials. Indeed, one avenue to searching for novel skyrmionics spin textures has been to explore magnetic materials with a certain crystal symmetry that is then reflected in the form of the DMI. This led, for example to the discovery of the anti-skyrmion in two distinct materials[19,20]. An alternative strategy is to lower the symmetry of a high symmetry material, for example, by the introduction of strain along one direction or by the introduction of a strain gradient e.g. by mechanical means[21] or by ripples in thin film membranes[22] or by thin film epitaxy[23]. Although strain gradients have been observed in various oxide thin films[24] giving rise to flexoelectricity[25] it has been proven to be much more difficult to establish strain gradients in simple metallic films[26,27]. Here we show vertical strain gradients over considerable distances in metallic films formed from cubic Co-Ni-Al alloys using an unusual underlayer of the $L1_0$ ordered alloy Ir-Al.

Typically, thin films deposited on a substrate, via single crystalline or polycrystalline epitaxy, maintain a constant strain up to a critical thickness when the strain is released. We show that thin films of the tetragonal ferromagnets $Co_{2.3}Al$ and $Co_{2.58}Ni_{0.26}Al$ rather display a significant strain gradient when deposited on a suitable underlayer and when their thickness lies within a certain range (30~50 nm). To achieve large strain gradients, an optimized underlayer formed from one of the M-Al alloys (M = Ir, Pd and Ru) is introduced. These cubic alloys are chosen because at ambient temperature they grow as flat films with a highly chemically ordered $L1_0$ structure[28,29] and are characterized by a larger lattice parameter than the ferromagnetic layers considered here. The largest strain gradient ($\nabla_t \varepsilon = 3 \times 10^{-3}$ Å/unit cell (uc)) was found for a ferromagnetic layer of $Co_{2.58}Ni_{0.26}Al$ on an IrAl underlayer. For strain gradients larger than approximately $5\sim6\times10^{-4}$ Å/uc, Néel skyrmions are directly observed by Lorentz Transmission Electron Microscopy (LTEM) in both $Co_{2.3}Al$ and $Co_{2.58}Ni_{0.26}Al$, in good agreement with *ab-initio* calculations of the strain gradient and the resulting DMI that we discuss below. For the case of $Co_{2.3}Al$, high temperature LTEM studies show clear evidence for skyrmions up to very high temperatures of ~770 K

The strain-gradient in the epitaxial films was determined by fitting the x-ray diffraction (XRD) reflection profiles collected by line scans along the $q_z$ direction in the vicinity of the (002) reflection in reciprocal space. The simulation of the reflection profiles was carried out by employing the recursive matrix formalism in the kinematic scattering approximation[30,31]. The film is sub-divided into thin slices, each having a different strain ($\varepsilon$) along the surface normal [001] as schematically shown in Fig. 1b. Here, and in the following, we refer to the *c*-lattice parameter of the cubic IrAl underlayer (2.985 Å). Additional fitting parameters include the film thickness $t$ and the root-mean squared (rms) surface roughness ($\sigma$) modelled by a Debye-Waller-type approach. The finite instrumental resolution along $q_z$ was taken into account by using a pseudo-Voigt function whose full width at half maximum ($\Delta q_z$) was derived from a scan across the Si (202) reflection ($\Delta q_z$ was found to be $\approx 2.2\times10^{-3}$ Å$^{-1}$ which corresponds to $6.5\times10^{-3}$ reciprocal lattice units). A least squares fit of the simulated curve to the experimental data was carried out using a standard $\chi^2$ minimization procedure[32]. Results for 30 and 50 nm thick $Co_{2.3}Al$ and 27 nm thick $Co_{2.58}Ni_{0.26}Al$ layers grown on top of IrAl are compared in Fig. 1c-e. The experimental (00L) reflection profiles (symbols) are shown together with the corresponding calculated ones (solid line) on a logarithmic intensity scale. The analysis indicates that these films exhibit a significant positive strain gradient which ranges from approximately $6.8 \times 10^{-4}$ Å/uc (Fig. 1c and d) up to $3 \times 10^{-3}$ Å/uc (Fig. 1e). As an example, Fig.1e shows details of the refined strain profile within the 50 nm thick $Co_{2.3}Al$ film. Next to the IrAl layer we find an approximately 9 nm thick $Co_{2.3}Al$ layer which is constantly strained by about -8.0% ($c$ = 2.74 Å) relative to the lattice parameter of IrAl of 2.985 Å. This layer is followed by an approximately 30 nm thick $Co_{2.3}Al$ layer in which the strain linearly diminishes from -8.0% at the bottom to -5.7% (2.81 Å) at the top corresponding to a vertical strain gradient of $6.8 \times 10^{-4}$ Å/uc. There is also a 10~20% discrepancy between the nominal film thickness and that derived by XRD. We believe that this can be attributed to the lack of well-defined long-range order in some part of the film structure. Similar results are found for all films in this thickness regime. Our findings show that films in the thickness regime well outside the 30~50 nm regime exhibit much smaller strain gradients ($\approx 2 \times 10^{-4}$ Å/uc) and, correspondingly, no Néel type skyrmions. Either the majority of the film volume is in a constantly strained state for $t$ < 20 nm or strain relaxation sets in for $t$ > 50 nm.

The accurate modelling of the strain profile, as outlined above, is a prerequisite to obtain a high degree of agreement between the simulated and experimental reflection curves. A large strain gradient is directly evidenced by the pronounced asymmetry of the x-ray reflection profile, which is well reproduced by the simulations (Fig. 1c-e). By contrast a strain gradient in the lower $10^{-4}$ Å/uc regime results in a nearly symmetric profile (see Fig. S12b, c, and d). In agreement with the (linear) positive strain gradient derived from the profile fitting we find that the average $c$ lattice parameter of the $Co_{2.3}Al$ films derived by the peak position increases with film thickness (see Fig. S12a). The high quality of the epitaxially grown $Co_{2.3}Al$ films on IrAl was directly proved by cross-sectional high-resolution transmission electron microscopy (HRTEM). A typical image is shown in Fig. 1f together with its Fourier transforms (FT) showing bright, well defined spots. Vibrating sample magnetometry (VSM) measurements confirmed the out-of-plane easy axis magnetization of IrAl|Co$_{2.3}$Al bilayers by recording hysteresis loops in the temperature range of 100-600 K: exemplary data for one sample are shown in Fig. S8. The magnetic moment gradually decreases with increasing temperature, while the hysteresis profile is retained up to ~600 K. The Curie temperature ($T_c$) is estimated to be ~1200 K (inset of Fig. S8) from fitting $M$ (T) to Bloch's law (see SI).

Figure 2a shows a schematic of the LTEM configuration in which a thin lamella, that has been prepared from a deposited film, is oriented so that its surface normal is parallel to the electron beam. This situation corresponds to the tilt angles $\alpha=\beta=0$ where these angles correspond to rotations about two mutually orthogonal rotation axes lying in the lamella plane (the sample holder allows maximum tilt angles of $\alpha = \pm 36°$ about the *x*-axis and $\beta = \pm 31°$ about the *y*-axis). The lamella is prepared by initially mechanically polishing the backside of the MgO substrate followed by argon ion milling until the 200 keV electron beam used in the LTEM experiments can penetrate the lamella. We have also prepared free-standing films for TEM imaging, as discussed later. First, we discuss results for a 4.3 nm IrAl|30 nm Co$_{2.3}$Al bilayer. No magnetic contrast is observed for $\alpha=\beta=0°$ (see Fig. 2b). However, a magnetic contrast appears, if the sample is tilted, and many circular objects are seen. Each of these objects exhibits bright and dark regions at their opposite edges. These regions appear along the *x(y)* axis when the lamella is tilted by $\pm \alpha(\beta)$ (Fig. 2c-f). This magnetic contrast is characteristic of a Néel-type skyrmion[33,34]. We find that these nano-objects are stable over a wide range of magnetic field at ambient temperature and to low temperatures (see Fig. S20 and S21). In addition to the observation of Néel skyrmions in Co$_{2.3}$Al, clear evidence of Néel-type skyrmions is also provided by LTEM for a 4.3 nm IrAl|27 nm Co$_{2.58}$Ni$_{0.26}$Al thick bilayer thin film, as shown in

Fig. 2g and h (for more information, refer to Fig. S24 and S25). For this film we derive the maximum strain gradient to be $\nabla_t\varepsilon \sim 3\times10^{-3}$ Å/uc.

Next, we discuss in detail five $Co_{2.3}Al$ samples with varying thickness of the ferromagnetic layer (See Fig. 3). A detailed field ($\mu_0H$) versus thickness ($t$) phase diagram showing the stability regions of the different spin textures is outlined in Fig. 3a. Samples s1, s2 and s5 exhibit reflection profiles that are nearly symmetric (Fig. S12) corresponding to $\nabla_t\varepsilon$ lying in the low $10^{-4}$ Å/uc regime (Fig. 3b). For these samples, Néel-type skyrmions cannot be stabilized and only stripe domains or type-II bubbles or both of them are observed (Fig. S21). By contrast, samples s3 and s4 have the largest strain-gradients (up to $6.5\times10^{-4}$ Å/uc) among these 5 samples (see Fig. 3b and S21) and Néel skyrmions are observed. We find, therefore, that Néel skyrmions are only observed for film thicknesses in the 30 to 50 nm regime and for intermediate magnetic field strengths. Thinner films exhibit stripe-type domains whereas thicker films exhibit both stripe domains[35] and type-II bubbles[36] (see Fig. S21). Apart from the formation of Néel skyrmions, the effect of the strain gradient also is clearly manifested in the cycloidal domains observed in the absence of a magnetic field, as discussed in Fig. 3c. Here, the size ($\lambda$) of the cycloidal domains displays a clear inverse relationship versus the strain gradient ($\nabla_t\varepsilon$) except for a large uncertainty in sample s2. The distinction between the cycloidal phase and stripe domains can be identified from tilting experiments in LTEM. The cycloidal phase, characterized by Néel-type walls, generates contrast only when the sample is tilted, whereas stripe domains with Bloch walls would produce contrast even without tilting.

We find that large strain gradients give rise to an increase in the skyrmion diameter ($d_{sk}$): for example, $d_{sk} = 315 \pm 23$ nm for 27 nm thick $Co_{2.58}Ni_{0.26}Al$ layers, whereas $d_{sk} = 95 \pm 7.5$ nm for 30 nm thick $Co_{2.3}Al$ layers (see Fig. 3d). This finding is supported by the larger ratio of $D/A$ for a larger strain gradient, where $A$ is the spin stiffness and $D$ is the DMI constant, as substantiated by *ab*-initio calculations (See Fig. 5b)[37].

Only type-II bubbles are found in $Co_{2.3}Al$ films grown without an underlayer, irrespective of thickness, as shown in Fig. S19. Furthermore, when the composition (*x*) of $Co_xAl$ is modified to *x* = 2.0, 2.6, and 2.9, or when IrAl is replaced by PdAl or RuAl, only stripe domains and type-II bubbles are found (see Fig. S22 and S23). In order to investigate the role of a possible interfacial DMI contribution between IrAl and $Co_{2.3}Al$ in stabilizing Néel-type skyrmions, two sets of samples were prepared that include, in one case, the insertion of a 0.3 nm thick aluminum dusting layer between 4.3 nm IrAl and 30 nm $Co_{2.3}Al$ and, in a second case, the addition of a 2 nm thick IrAl layer on top of the 4.3 nm IrAl|30 nm $Co_{2.3}Al$ bilayer structure. Néel skyrmions were observed in both cases but their diameter was unchanged (see LTEM images in Fig. S27 and S28) showing that the role of any interfacial DMI is negligible as compared to the strain gradient effect (See Fig. S26 and S27).

To explore the thermal stability of the Néel skyrmions, in-situ LTEM experiments were carried out up for temperatures as high as 770 K. A novel method was developed to make this experiment possible. A freestanding membrane formed from 12 nm MgO|4.3 nm IrAl|30 nm $Co_{2.3}Al$ was prepared by first growing this structure on a sacrificial layer of $Sr_3Al_2O_6$ that had been deposited by pulsed laser deposition on a $SrTiO_3$(001) substrate. The $Sr_3Al_2O_6$ layer was subsequently dissolved in water from the edges of the substrate and the membrane floated off[38] and subsequently transferred to a heating chip (DENS Solutions), as shown in Fig. 4a (see Methods for more details). The chip was placed on a sample holder that allows for *in-situ* heating[39]. The membrane was then heated in steps of 100 K to 770 K at a heating rate of 5 K/sec. From selected area electron diffraction (SAED) the structure of the membrane was found to be highly thermally stable (see Fig. S32). LTEM images were recorded at each temperature using the same protocol. The temperature was increased in zero field and zero tilt. After the temperature was stabilized the sample was tilted (to *α*=20°) and then the field was increased in steps of 25 mT and a LTEM image taken at each field until the sample was fully field polarized.

A transition from a cycloidal domain state into a Néel skyrmion state with increasing magnetic field and finally a fully polarized state was found at all temperatures (see Fig. 4b-e). Both the skyrmion diameter and the skyrmion density decrease with increasing temperature (see Fig. 4e) which is presumably related to the decrease in saturation magnetization ($M_s$) with increasing temperature (see Fig. S8 and Fig. S29).

Next, we use first principles calculations to compute the magnitude of the DMI as a function of strain gradient. Based on a linear strain gradient scenario the generalized Heisenberg exchange tensor $J_{ij}$ was obtained from first-principles using the generalized magnetic force theorem[40-42], as implemented within a multiple scattering theory[43,44]. The diagonal elements of the exchange tensor represent the Heisenberg exchange coupling parameters $J_{ij}$ while a superposition of off-diagonal tensor elements corresponds to the DMI parameters, $D_{ij}$. $Co_{2.3}Al$ and $Co_{2.58}Ni_{0.26}Al$ films of various thicknesses were considered on a semi-infinite IrAl(001) substrate terminated by a semi-infinite CoAl bulk. To avoid strong perturbation effects induced by the IrAl substrate, the spin-orbit interaction in Ir was switched off. Disorder effects were taken into account within a coherent potential approximation[45,46].

A linear strain gradient is considered by assuming an increase of the $c$-lattice parameter beginning with $c_1 = 2.7$ Å for the first unit cell (uc) at the $Co_{2.3}Al$ ($Co_{2.58}Ni_{0.26}Al$) interface with IrAl up to $c_n = 2.933$ Å for the $n$-th unit cell at the top surface of the layer. The latter value corresponds to the limit where the topmost unit cell becomes cubic given the in-plane film lattice parameter of 2.933 Å. Thus, the lattice parameter $c_i$ of the $i$-th unit cell from the interface is given by $c_i = c_1 + \nabla_t\varepsilon \times (i-1)$, where $\nabla_t\varepsilon$ denotes the strain gradient in Å/uc. For $\nabla_t\varepsilon =$ 0.024 Å/uc, 0.0116 Å/uc, 0.006 Å/uc, 0.005 Å/uc and 0.003 Å/uc the $Co_{2.3}Al$ and $Co_{2.58}Ni_{0.26}Al$ films consist of 10, 20, 45, 60, and 80 uc's, respectively. The lowest strain gradient corresponds to the largest experimental value found in $Co_{2.58}Ni_{0.26}Al$. Smaller strain gradients were not considered because of the very large computational demand. The calculations describe well the strain induced emergence of the DMI.

Fig. 5a and b show the calculated $D/A$ ratio as a function of strain gradient in $Co_{2.3}Al$ and $Co_{2.58}Ni_{0.26}Al$, respectively. Here $A$ is the spin stiffness and $D$ is the DMI constant as conventionally used in micromagnetic simulations. The parameters $A$ and $D$ are defined in detail in the Supplementary Information. These values were obtained from the calculated atomistic $\boldsymbol{J_{ij}}$ and $\boldsymbol{D_{ij}}$ parameters using a conversion formula similar to that outlined in ref. 47. In general, $D/A$ is proportional to the strain gradient. While the spin stiffness $A$ is almost constant, the DMI parameter $D$ increases monotonically. For the case of $Co_{2.58}Ni_{0.26}Al$, both $A$ and $D$ are smaller than in $Co_{2.3}Al$, since the presence of Ni reduces the magnetic interaction in CoAl alloys. Nevertheless, the $D/A$ ratio, which is mainly responsible for skyrmion formation, is calculated to be similar in both systems for similar strain gradients. Experimentally, we find a much larger strain gradient for $Co_{2.58}Ni_{0.26}Al$ as compared to $Co_{2.3}Al$ and, correspondingly, as expected, a larger skyrmion diameter, as discussed above (Fig. 3d).

Real space images were simulated by micromagnetic simulations using the Mumax3 solver[46]. Material parameters were partially taken from experiment (saturation magnetization, uniaxial anisotropy) and partially from calculations ($A$, $D$). In agreement with the LTEM images (see Fig. 2), the micromagnetic simulations show a cycloidal magnetic ground state in the absence of an external magnetic field, which is shown in Fig. 5c. Applying a perpendicular magnetic field results in the formation of magnetic Néel skyrmions, which can be distinguished by determining their topological charge (See Fig. S33) [2]. Further increase of the external magnetic field results in a fully polarized state in agreement with experiment.

In summary, we have demonstrated the presence of chiral Néel-type spin textures in thin films of simple metallic ferromagnetic alloys, that have high Curie temperatures, at elevated temperatures of up to ~770 K. These spin textures arise from the creation of a large strain gradient throughout the thickness of the film via thin film epitaxy. The strain gradient lowers the symmetry of the centrosymmetric magnetic material and, thereby, gives rise to a DMI within the interior of the film. Our findings significantly broaden the range of potential magnetic materials capable of exhibiting complex chiral spin textures, which are highly relevant for numerous spintronic applications. Thus, our approach to generating "hot" skyrmions can be extended to other magnetic systems with high magnetic ordering temperatures.

## Methods

**Thin films growth, characterization**

The films were deposited in an AJA 'Flagship Series' sputtering system in the presence of Ar gas on 10×10 mm² MgO substrates with [001] orientation. The base pressure before deposition was less than $10^{-8}$ Torr and the pressure during deposition was 3 mTorr. The IrAl, PdAl, RuAl Co$_x$Al and Co-Ni-Al, alloy thin films were prepared by co-sputtering from individual heavy metal and Aluminum targets with 2-inch diameter and 0.25-inch thickness. The composition of thin films is calibrated by non-destructive Rutherford Backscattering spectroscopy. The atomic ratio of IrAl, PdAl, and RuAl is 42:58, 35:65 and 46:54 respectively. The highly resistive TaN-capping layer was prepared by introducing 20% $N_2$ into the Ar gas flow. The magnetization hysteresis loops were measured with a superconducting quantum interference device vibrating sample magnetometer (SQUID-VSM).

**X-ray analysis of the strain gradient in thin MAl (M= Ru, Pd and Ir)|Co$_x$Al (Co-Ni-Al) films**

The magnitude of the strain gradient, $\nabla_t \varepsilon = \partial \varepsilon_t / \partial t$, along the sample normal ($c$-axis) was analyzed by X-ray diffraction probing the reflection profiles collected by line scans in reciprocal space along the q$_z$ direction in the vicinity of the (002) reflection. The experiments were carried out in our home laboratory using a Ga-jet x-ray source (λ=1.3414 Å) and a six-circle diffractometer as well as at the beamline 25b of the European Synchrotron Radiation Source (ESRF) in Grenoble (France). In both cases, a two-dimensional pixel detector was used employing a region of interest to provide high resolution in k-space along the longitudinal scan direction (q$_z$).

**Magneto-transport measurements**

The Hall resistivity was measured using the van der Pauw method using 3×3 mm² square samples cut from 10×10 mm² blanket films. A direct current source (Keithley 6221), and a nanovoltmeter (Keithley 2182a) were used for the electrical measurements.

**Transmission electron microscopy**

For the transmission electron microscopy (TEM) investigations, cross sectional lamellae from the thin films were prepared by Focused Ion Beam (FIB) Ga+ ion milling [TESCAN GAIA3 operating at 30 kV ion-beam energy] using standard lift-out procedures. For Lorentz TEM imaging several plane-view lamellae were prepared from the as-deposited films by Ar ion milling from the back-side of the MgO substrate. Structural imaging was performed using JEOL ARM300F2 TEM. Magnetic textures were investigated by TEM [FEI TITAN 80-300 and JEOL JEM-F200] in the Lorentz mode operated at an accelerating voltage of 300 kV and 200 kV respectively, using a GATAN double-tilt LN2 cooling holder (Model 636) which is capable of varying the temperature. A vertical magnetic field was applied to the lamella within the TEM column by passing currents through the coils of the objective lens and a Lorentz mini-lens was used for imaging.

**Free-standing thin film transfer procedure**

For the LTEM investigations on samples at various temperatures, a lift-off and transfer method was used. To protect the entire structure, a 100 nm thick PMMA layer was coated onto the as deposited sample that was prepared by a combination of pulsed laser deposition and sputter deposition in two separate chambers (Fig. 4a). The multilayered structure was immersed in de-ionized water for ~30 min to remove the $Sr_3Al_2O_6$ layer (Fig. 4a). The separated sheet together with water was then picked up and transferred onto the heating chip. Before the LTEM measurements, the protective PMMA layer was removed by oxygen plasma after drying in nitrogen for 6 hours.

**DFT calculations**

First-principles calculations were performed using a self-consistent fully relativistic Green function method[43,44], which is specially designed for semi-infinite systems such as surfaces and interfaces[47]. A generalized gradient approximation was utilized for the exchange-correlation potential[46] and disorder effects were taken into account within a coherent potential approximation as it is implemented in the multiple scattering theory[45]. The generalized Heisenberg exchange tensor was estimated using the magnetic force theorem[40] adopted for the relativistic case[41,42]. The Curie temperature of $Co_xAl$ alloys was calculated within a random phase approximation[2]. More details about DFT calculation, see supplementary information.

**Data availability**

The data that support the findings of this study are available from the corresponding author upon on reasonable request.

**Acknowledgements**

We acknowledge funding from the Deutsche Forschungsgemeinschaft (DFG, German Research Foundation) – Project number 403505322 under SPP 2137. The authors thank Claudia Münx, and Norbert Schammelt for preparing plane-view and cross-sectional TEM lamellae. The authors acknowledge the help of Katayoon Mohseni for the XRD data analysis. We thank Herbert Engelhard for conducting Rutherford Backscattering measurements of the composition of the thin films. We would like to thank Ankit K. Sharma for support in the MFM imaging. A.E. acknowledges funding by the Fonds zur Förderung der Wissenschaftlichen Forschung (FWF) under Grant No. I 5384. Calculations were carried out at the Rechenzentrum Garching of the Max-Planck Society. This theory calculation was supported by hardware grants (PI Kramer, JKU Linz) from NVDIA and utilized NVDIA A100 GPUs. The authors thank the European Synchrotron Radiation Facility (ESRF) for provision of beamtime.

**Author contributions**

S.S.P.P. and P.W. initialized the project. P.W. optimized all thin film stacks, carried out thin film growth, characterization, and magneto-transport measurements. R.S. and H.D. performed in-situ LTEM experiments and R.S. analyzed the LTEM images. H.L.M, J.R.-Z. and E.S.-T. carried out the XRD measurements of the films. K.G. transferred the freestanding films into heating chips and $Si_3N_4$ membranes. A.M. conducted high-temperature VSM measurements.

I.K. conducted RBS analysis of the thin film samples. A.E. and D.E. carried out first principles calculations and micromagnetic simulations. H.D. carried out HRTEM imaging. P.W., R.S., H.L.M., A.E., and S.S.P.P wrote the manuscript. All authors discussed the data and commented on the manuscript. P.W. and R.S. contributed equally to this work. S.S.P.P. supervised the project.

**Competing interests**

The authors declare no competing financial interests. Correspondence and requests for materials should be addressed to S.S.P.P. (stuart.parkin@mpi-halle.mpg.de).

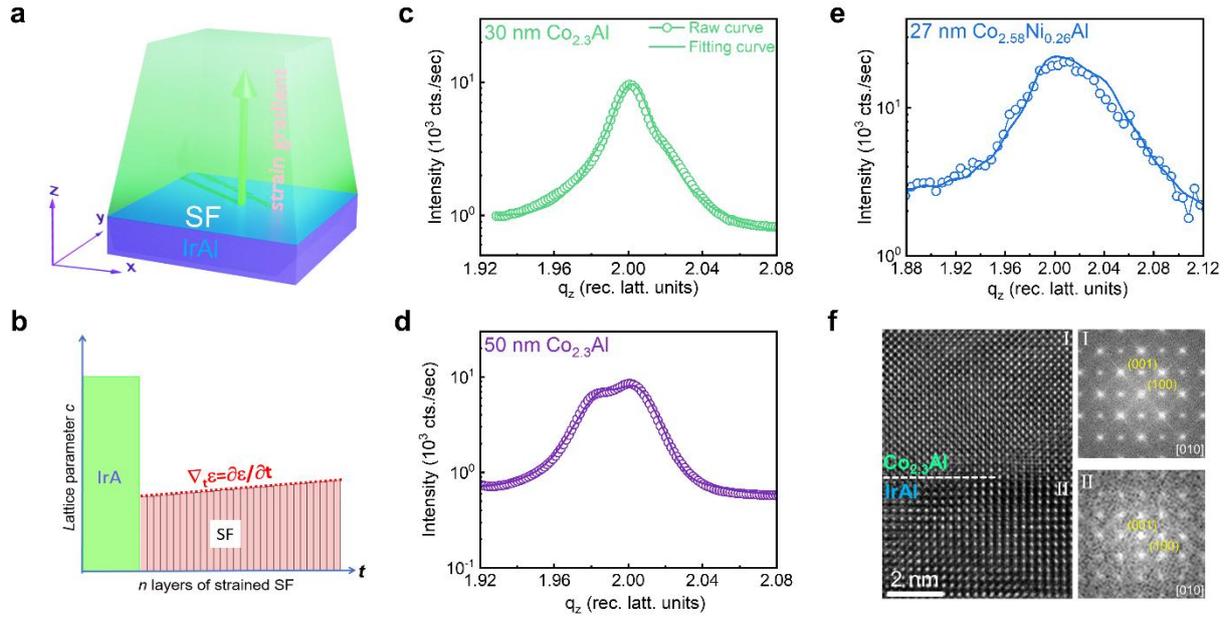

**Fig. 1: Strain gradient analysis and structural properties. a**, Schematic illustration of the epitaxial relationship between a strained ferromagnetic film (SF) on an IrAl(001) underlayer. **b**, Schematic of the XRD derived strain profile of a $Co_{2.3}Al$ film on IrAl(001). A constantly strained layer ($\nabla_t\varepsilon = 0$) is located between the IrAl layer and the SF which is characterized by a linear positive strain gradient. **c-e**, Experimental (circles) and calculated (solid line) x-ray diffraction profiles on a log scale in the vicinity of the L=2 reflection along the [00L] direction in reciprocal space. Profiles of 30 nm ($\nabla_t\varepsilon = 6.8\times10^{-4}$ Å/uc), 50 nm ($\nabla_t\varepsilon = 5.8\times10^{-4}$ Å/uc) $Co_{2.3}Al$ and 27 nm thick $Co_{2.58}Ni_{0.26}Al$ films ($\nabla_t\varepsilon = 3.0\times10^{-3}$ Å/uc) exhibit a pronounced asymmetry directly indicating a large strain gradient. **f**, Cross-sectional, bright field, high-resolution transmission electronic microscopy (HR-TEM) image of an IrAl|$Co_{2.3}Al$ bilayer. Corresponding Fourier transform (FT) of two different regions (I, and II) showing the epitaxial relationship of the IrAl and $Co_{2.3}Al$ layer.

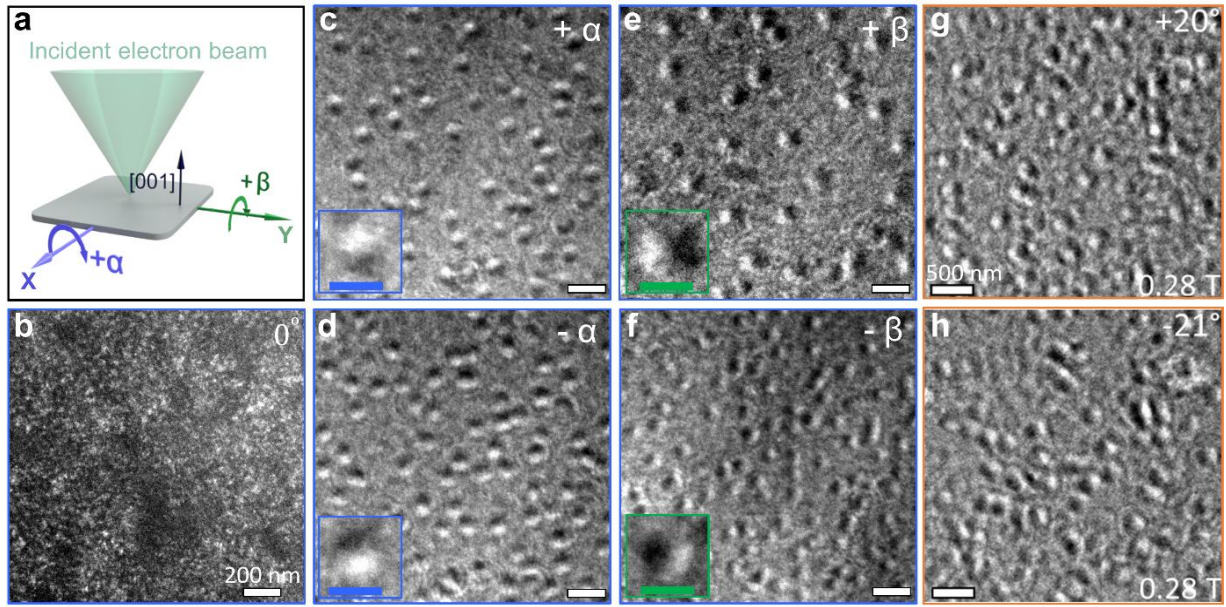

**Fig. 2: Lorentz transmission electron microscopy (LTEM) images of free-standing films at 300 K.** **a**, Schematic illustration of specimen tilting inside the electron microscope. Zero tilt corresponds to the incident electron beam being oriented along the surface normal [001] of the lamella. **b-f**, LTEM images recorded for a 4.3 nm IrAl|30 nm $Co_{2.3}$Al bilayer in the presence of a 0.25 T magnetic field applied along the TEM column. (**b**) LTEM image at zero tilt shows no magnetic contrast. **c-f**, LTEM images recorded upon sample tilting about the *x* axis (**c** and **d**) $\alpha = \pm30°$, and the *y* axis (**e** and **f**) $\beta = \pm30°$. The insets in **c-f** show magnified images of the highlighted (blue or green circle) nano-object that corresponds to a Néel skyrmion. All images are taken at a defocus value of -1.5 mm. The scale bar for **b-f** is 200 nm and the scale bar inset **c-f** is 100 nm. **g**, **h**, LTEM images of bilayer of 4.3 nm IrAl|27 nm $Co_{2.58}Ni_{0.26}$Al in the presence of a 0.28 T magnetic field applied along the TEM column at 300 K. Defocus was set to -0.6 mm, tilt angle ($\alpha$) to +20° and -21°, the scale bar for **g** and **h** is 500 nm.

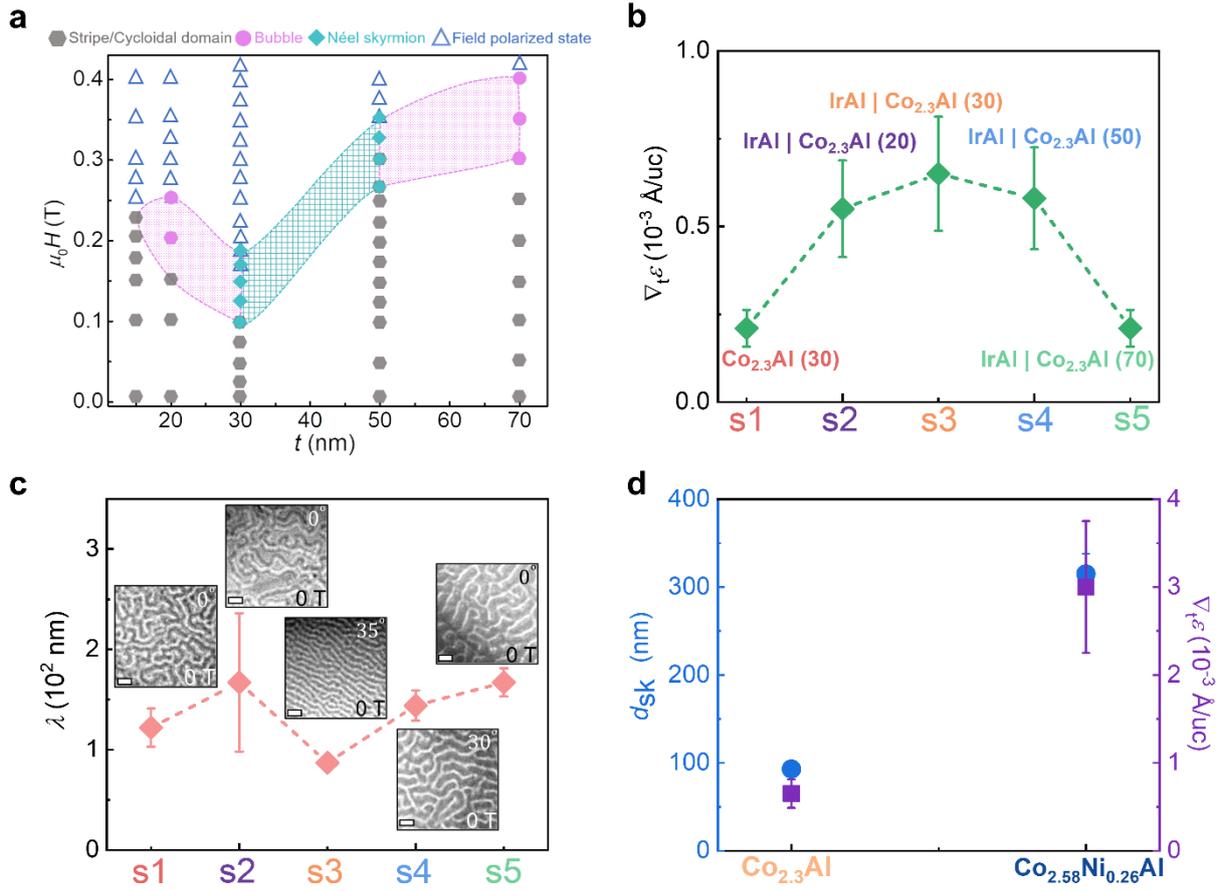

**Fig. 3: Thickness-dependent magnetic textures and strain gradient at ambient temperature. a**, Dependence of the spin texture type derived from LTEM imaging for IrAl|Co$_{2.3}$Al ($t$) bilayers as a function of the Co$_{2.3}$Al thickness and applied magnetic field. Different spin textures are labelled by: stripe/cycloidal: ⬢, bubbles: ●, Néel skyrmion: ◆, and △:field polarized state. **b,c**, show data for 5 samples: **s1**: 30 nm Co$_{2.3}$Al; **s2**: 4.3 nm IrAl|20 nm Co$_{2.3}$Al; **s3**: 4.3 nm IrAl|30 nm Co$_{2.3}$Al; **s4**: 4.3 nm IrAl|50 nm Co$_{2.3}$Al; **s5**: 4.3 nm IrAl|70 nm Co$_{2.3}$Al. (**b**) Plot of measured strain gradient and (**c**) variation of period of stripe/cycloidal versus sample number. The insets show the corresponding LTEM images of stripe/cycloidal domains in zero magnetic field. **d**, Dependence of the skyrmion diameter and strain gradient in 4.3 nm IrAl|30 nm Co$_{2.3}$Al and 4.3 nm IrAl|27 nm Co$_{2.58}$Ni$_{0.26}$Al bilayer thin films. In this figure all the lamellae were prepared by conventional Ar ion milling.

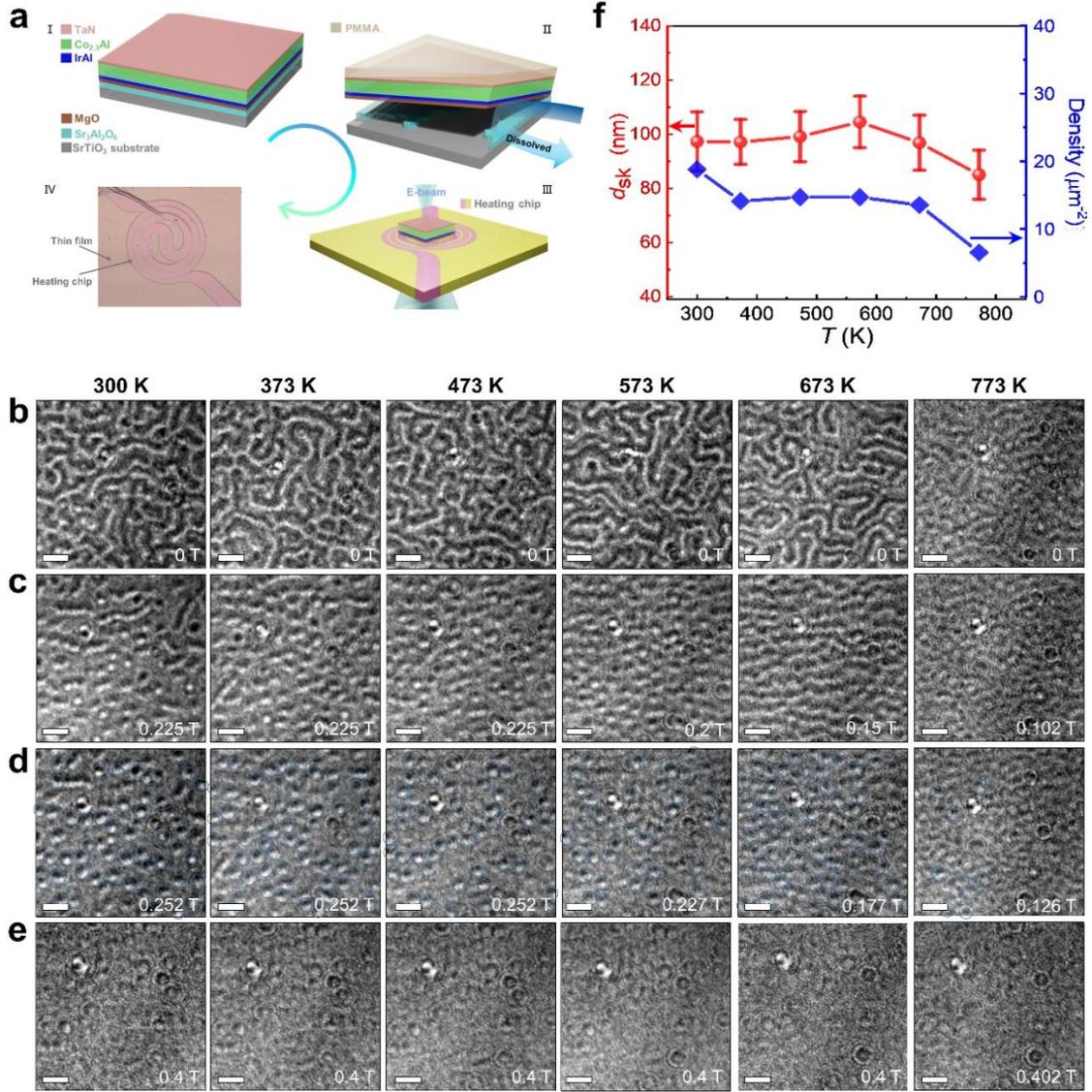

**Fig. 4: Lorentz transmission electron microscopy (LTEM) images of a freestanding 4.3 nm IrAl | 30 nm Co$_{2.3}$Al bilayer as a function of temperature. a**, Schematic diagram of the formation and transfer of a MgO|4.3 nm IrAl|30 nm Co$_{2.3}$Al|TaN membrane onto a heating chip for in-situ LTEM measurements. (I) Schematic of a MgO|IrAl|Co$_{2.3}$Al|TaN thin film heterostructure deposited onto a Sr$_3$Al$_2$O$_6$ (SAO) buffer layer grown by pulsed laser deposition on a SrTiO$_3$ (001) substrate. (II) Separation of the multilayer film in deionized water by completely dissolving the SAO layer after spin-coating a PMMA resist onto the TaN surface. (III) Transfer of the thin film onto the heating chip (Wildfire, DENS). (IV) Optical microscopy image of the membrane on the heating chip. **b-e**, LTEM images recorded at temperatures between 300 and 770 K at selected magnetic fields that are given on each image. LTEM images of (**b**) cycloidal phase, (**c**) mixture phase of magnetic cycloidal phase and Néel-type skyrmions, (**d**) Néel-type skyrmions with blue circles and (**e**) field polarized states. **f**, Variation of skyrmion diameter (red circles) and the skyrmion density (blue squares) as a function of temperature: the error bars represent the standard deviation in the skyrmion diameter. All LTEM images are recorded at a 20º tilt at a defocus value of -1 mm. The scale bar in each LTEM image is 200 nm.

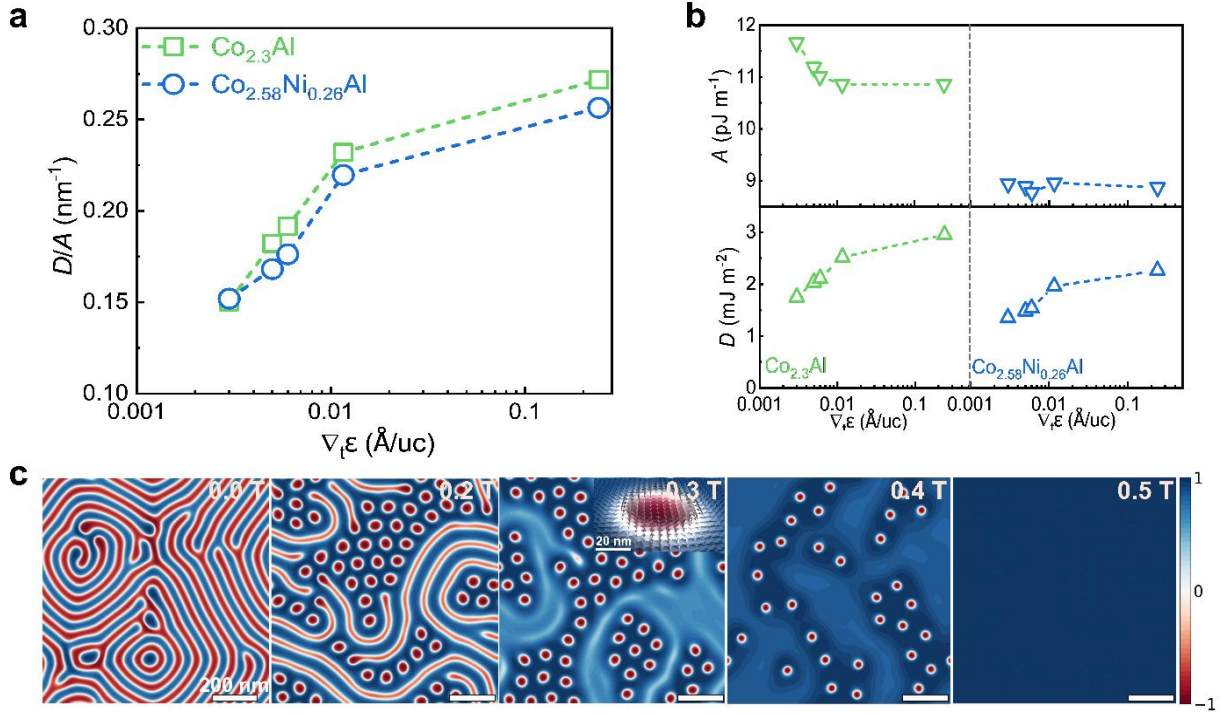

**Fig. 5: First-principles calculations of the generalized exchange interaction parameters and micromagnetic calculations. a,b,** Ratio $D/A$ versus strain gradient ($\nabla_t\varepsilon$) for $Co_{2.3}Al$ and $Co_{2.58}Ni_{0.26}Al$, (**b**) show parameters $D$ and $A$ versus strain gradient. **c**, Micromagnetic simulations of the spin textures for $\nabla_t\varepsilon = 3\times10^{-3}$ Å/uc varied as a function of an applied magnetic field (see labels) along the surface normal. At $\mu_0H_z = 0$ T a cycloidal state is observed. Increasing the magnetic field first leads to a mixture of magnetic stripes and Néel skyrmions then followed by Néel skyrmions and gradually into a fully polarized state (from left to right). The inset at $\mu_0H_z = 0.3$ T shows a close-up view of one Néel skyrmion. The color code depicts the normalized projection of the magnetization onto the surface normal, with +1, 0, and -1 representing up, in plane, and down, respectively.